\documentclass[letterpaper,num-refs,dvipsnames,svgnames]{oup-contemporary}
\journal{gigascience}

\usepackage{amsmath,amssymb}
\usepackage{capt-of}
\usepackage{cleveref}
\usepackage{graphicx}
\usepackage{xspace}

\usepackage{natbib}

\usepackage{ifthen}
\newboolean{draftmode}
\setboolean{draftmode}{true}
\ifthenelse{\boolean{draftmode}}{%
  \newcounter{comments}
  \newcommand{\jeff}[1]{\addtocounter{comments}{1}{\color{orange}[Jeff \thecomments: #1]}}
  \newcommand{\todojeff}[1]{\addtocounter{comments}{1}{\color{orange}[For Jeff \thecomments: #1]}}
  \newcommand{\tmm}[1]{\addtocounter{comments}{1}{\color{ForestGreen}[TMM \thecomments: #1]}}
    
  \newcommand{\markc}[1]{\addtocounter{comments}{1}{\color{Mahogany}[Mark \thecomments: #1]}}  
 \newcommand{\todomarkc}[1]{\addtocounter{comments}{1}{\color{Mahogany}[For Mark \thecomments: #1]}}  
  \newcommand{\meghana}[1]{\addtocounter{comments}{1}{\color{TealBlue}[Meghana \thecomments: #1]}}

   \newcommand{\kyle}[1]{\addtocounter{comments}{1}{\color{NavyBlue}[Kyle \thecomments: #1]}}  
  \newcommand{\todokyle}[1]{\addtocounter{comments}{1}{\color{NavyBlue}[For Kyle \thecomments: #1]}}  
   \newcommand{\nure}[1]{\addtocounter{comments}{1}{\color{CornflowerBlue}[Nure \thecomments: #1]}}  
   
 \newcommand{\todonure}[1]{\addtocounter{comments}{1}{\color{CornflowerBlue}[For Nure \thecomments: #1]}}

  \newcommand{\new}[1]{{\color{brown}#1}}
\newcommand{\forversiontwo}[1]{{\color{RubineRed}[\textbf{V2:} #1]}}
\newcommand{\forversionthree}[1]{}
}{
\newcommand{\jeff}[1]{}
\newcommand{\todojeff}[1]{}
\newcommand{\tmm}[1]{}
\newcommand{\kyle}[1]{}
\newcommand{\todokyle}[1]{}
\newcommand{\markc}[1]{}
\newcommand{\todomarkc}[1]{}
\newcommand{\meghana}[1]{}
\newcommand{\nure}[1]{}
\newcommand{\todonure}[1]{}
\newcommand{\new}[1]{{#1}}
\newcommand{\forversiontwo}[1]{}
\newcommand{\forversionthree}[1]{}
}

\newcommand{\drop}[1]{}

\newcommand{\pval}{\textit{p}-value\xspace}

\newcommand{\pvalue}[2]{\ensuremath{#1 \times 10^{#2}}}

\newcommand{\sars}{SARS-CoV\xspace}
\newcommand{\sarstwo}{SARS-CoV-2\xspace}
\newcommand{\covid}{COVID-19\xspace}

\newcommand{\sinksource}{SinkSource\xspace}

\newcommand{\genemania}{GeneMANIA\xspace}

\newcommand{\deepnf}{deepNF\xspace}
\newcommand{\rwr}{RWR\xspace}

\newcommand{\local}{Local\xspace}
\newcommand{\reals}{\ensuremath{\mathbb{R}}\xspace}

\title{Interpretable Network Propagation with Application to Expanding the Repertoire of Human Proteins that 
Interact with SARS-CoV-2} 
\author[1,\authfn{1}]{Jeffrey N. Law}
\author[1]{Kyle Akers}
\author[2]{Nure Tasnina}
\author[3]{Catherine M. Della Santina} 
\author[4]{Shay Deutsch}
\author[5]{Meghana Kshirsagar}
\author[6]{Judith Klein-Seetharaman}
\author[7]{Mark Crovella}
\author[8]{Padmavathy Rajagopalan}
\author[3]{Simon Kasif}
\author[2,\authfn{2}]{T. M. Murali}
\affil[1]{Interdisciplinary Ph.D. Program in Genetics, Bioinformatics, and Computational Biology, Blacksburg, VA, USA}
\affil[2]{Department of Computer Science, Virginia Tech, Blacksburg, VA, USA}
\affil[3]{Department of Biomedical Engineering, Boston University, Boston, MA, USA}
\affil[4]{Department of Mathematics, University of California, Los Angeles, CA, USA}
\affil[5]{AI for Good Lab, Microsoft, Redmond, WA, USA}
\affil[6]{Department of Chemistry, Colorado School of Mines, Golden, CO USA}
\affil[7]{Department of Computer Science, Boston University, Boston, MA, USA}
\affil[8]{Department of Chemical Engineering, Virginia Tech, Blacksburg, VA, USA}
\authnote{  \authfn{1} Current Address: National Renewable Energy Laboratory, Golden, CO, USA
\authfn{2} Corresponding author: murali@cs.vt.edu, 
ORCIDs: T. M. Murali [0000-0003-3688-4672];
Jeffrey N Law [0000-0003-2828-1273];
Kyle Akers [0000-0001-5093-4619];
Catherine M Della-Santina [0000-0002-6693-1022];
Shay Deutsch [0000-0001-7623-6376];
Meghana Kshirsagar [0000-0002-4673-614X];
Judith Klein-Seetharaman [0000-0002-4892-6828];
Mark Crovella [0000-0002-5005-7019];
Padmavathy Rajagopalan [0000-0001-9997-4953];
Simon Kasif [0000-0003-3297-9914];
}
\date{}

\begin{document}


\begin{frontmatter}
\maketitle


\begin{abstract}
\textbf{Background:} Network propagation has been widely used for nearly 20 years to predict gene functions and phenotypes. Despite the popularity of this approach, little attention has been paid to the question of provenance tracing in this context, e.g., determining how much any experimental observation in the input contributes to the score of every prediction. 
\textbf{Results:} We design a network propagation framework with two novel components and apply it to predict human proteins that directly or indirectly interact with SARS-CoV-2 proteins. First, we trace the provenance of each prediction to its experimentally validated sources, which in our case are human proteins experimentally determined to interact with viral proteins.
Second, we 
design a technique that helps to reduce the manual adjustment of parameters by users. We find that for every top-ranking prediction, the highest contribution to its score arises from a direct neighbor in a human protein-protein interaction network. We further analyze these results to develop functional insights on \sarstwo that expand on known biology such as the connection between endoplasmic reticulum stress, HSPA5, and anti-clotting agents. 
\textbf{Conclusions:}   
We examine how our provenance tracing method can be generalized to a broad class of network-based algorithms.
We provide a useful resource for the \sarstwo community that implicates many previously undocumented proteins with putative functional relationships to viral infection. This resource includes potential drugs that can be opportunistically repositioned  to target these proteins. We also discuss how our overall framework can be extended to other, newly-emerging viruses.

\end{abstract}

\end{frontmatter}

\drop{
\begin{keypoints*}
\todokyle{Choose key points}
\begin{itemize}
\item This is the first point
\item This is the second point
\item One last point.
\end{itemize}
\end{keypoints*}
}

\section{Background}
Network propagation algorithms have been widely used for nearly 20 years for function and phenotype prediction in systems biology~\citep{VFM+03,LK03,karaoz-kasif-whole-genome-annotation-pnas-2004,DCS04,fm-pvgf-2004,mwk-agfp-2006, is-pnd-2008}. More recently, applications of these techniques have included determination of genes associated with cancers and complex diseases~\citep{leiserson-raphael-hotnet2-natgen2015} and denoising single-cell gene expression data~\cite{2018-cell-van-dijk-peer-data-diffusion}. Nowadays, network-based algorithms facilitate  large-scale and automated data analysis of such complexity that it can be difficult for humans to understand the rationale that underlies a prediction, leading to decreased transparency and interpretability. 
%
%

In this work, we consider the fundamental problem of tracing the provenance of a prediction back to the experimental sources~\citep{2020-plos-bio-kasif-roberts-reproducible-trace}. Given a protein interaction network and a set of ``sources'', e.g., the human proteins that physically interact with \sarstwo~\citep{2020-nature-gordon-krogan-sars-cov-2-human-ppis}, suppose we apply a network-based algorithm to score and prioritize additional proteins that may directly or indirectly interact with the virus. Can we determine which source proteins make the highest contribution to the score computed for each prediction? Surprisingly, this question has been insufficiently studied in the field of network biology~\citep{2020-plos-bio-kasif-roberts-reproducible-trace}. This aspect takes particular importance in the context of \covid or other clinically or scientifically critical applications, where it may be important to understand the rationale behind the computational prediction of a new drug target before committing to expensive experimental validation.

We present a simple and direct method to solve this problem for a large class of network propagation algorithms. Specifically, for each protein $u$ in the network, we compute the precise contribution of each source to the score of $u$. This calculation enables us to sort the sources by their relative contributions to $u$ and to quantify the relative roles of sources at different distances from $u$.

\begin{figure*}
    \centering
    \includegraphics[width=\textwidth]{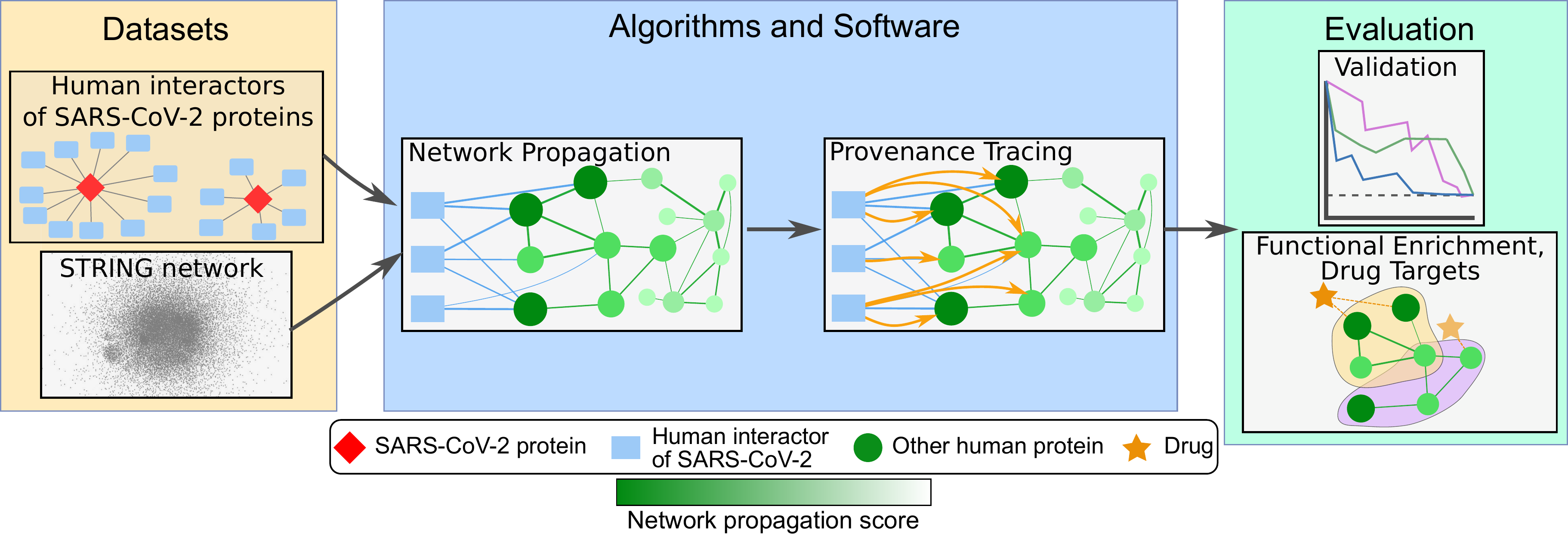}
    \caption{Overview of methodology. Algorithms and software for network propagation and provenance analysis take as input  experimentally determined host-pathogen protein interactions and a human protein interaction network. Evaluation includes cross-validation, functional enrichment, and literature-based examination of promising protein targets and drugs. 
    }
    \label{fig:prediction-pipeline}
\end{figure*}

To evaluate the effectiveness of this strategy, we apply it
to prioritize host proteins that may ``functionally'' (directly or indirectly) interact with SARS-CoV-2 proteins and host cellular processes that may be hijacked by the virus (\Cref{fig:prediction-pipeline}). 
To this end, we take advantage of a recently published dataset of human proteins that physically interact with \sarstwo~\citep{2020-nature-gordon-krogan-sars-cov-2-human-ppis}. Although these \sarstwo interactors are entry points to host cellular processes that may be hijacked by viral infection, the proteomics pipeline
used to discover them~\citep{2020-nature-gordon-krogan-sars-cov-2-human-ppis} may not capture \emph{in vivo} conditions  and  tissue-specific  interactions, leading to false negatives. Therefore, we apply network propagation algorithms 
to these known human protein interactors of SARS-CoV-2 proteins (sources) and a whole-genome human protein interaction network from the STRING database~\citep{szklarczyk-mering-string-v10.5-nar-2017}.
We identify statistically-enriched host biological processes and pathways that include highly-ranking proteins computed by our methods. We illustrate how our provenance analysis can simplify visualizations of these processes and assist in understanding how they  may be impacted by \sarstwo. 

\subsection{Data Description} 
\label{sec:datasets}
Here, we detail the different viral-human and human protein and functional interaction networks that we used in our study.

\paragraph{SARS-CoV-2--Human Protein-Protein Interactions (PPIs).} We obtained 332 human proteins that interact with SARS-CoV-2~\citep{2020-nature-gordon-krogan-sars-cov-2-human-ppis} and treated them as positive examples for our analysis. We added the ACE2 receptor to this set.

\begin{figure*}[htb]
    \centering
    \includegraphics[width=\textwidth]
    {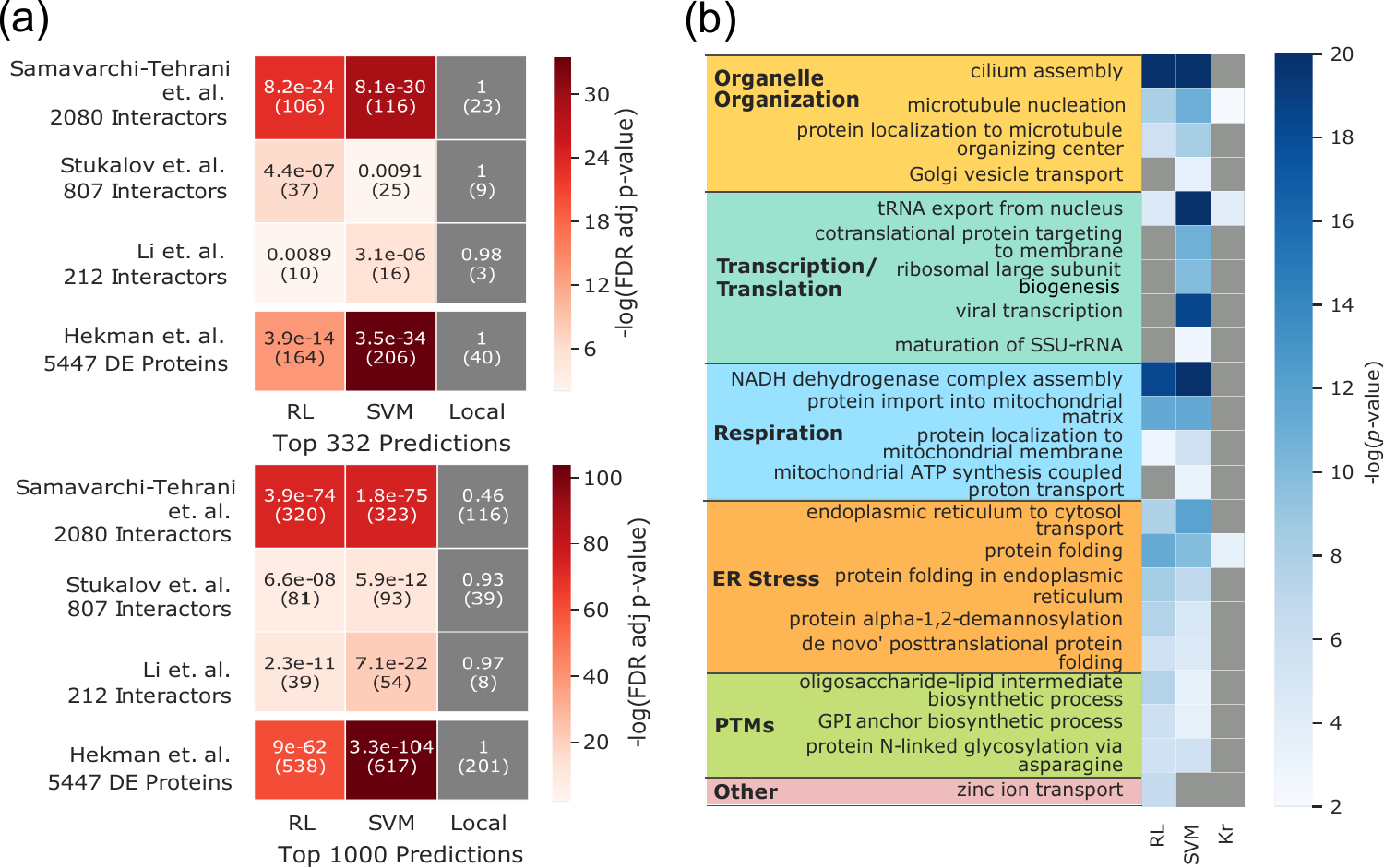}
    \caption{Network propagation results. 
    (a) Heatmap showing the FDR adjusted \pval from the hypergeometric test for the overlap between the top-ranking predictions of RL, SVM, and Local and three new experimental datasets of \sarstwo--human protein interactions~\citep{2020-biorxiv-samavarchi-tehrani-host-proximity-interactome,2020-biorxiv-li-guo-virus-host-interactome,2020-biorxiv-stukalov-multi-level-proteomics} and one dataset of differentially expressed (DE) proteins after \sarstwo infection~\citep{2020-mol-cell-hekman-sars-cov-2-host-responses}.  Each cell displays the FDR-adjusted \pval and the number of overlapping proteins in parentheses.  A gray cell indicates a \pval larger than 0.01.
  (b) Heatmap summarizing GO biological process terms enriched in top ranking proteins from RL and SVM
  and human interactors of \sarstwo proteins (indicated as `Kr'). We manually grouped the terms into broader categories shown in bold text. A gray cell indicates a \pval larger than 0.01. We examine the relevance of these biological processes to \sarstwo and \covid in ``Enriched Biological Processes'' in the supplementary results and in ``Discussion''.
    }
    \label{fig:prediction-results}
\end{figure*}

\drop{
\paragraph{Virus--Human PPIs for network proximity computations.} We obtained 104, 283, and 296 human proteins that interact with SARS-CoV, human immunodeficiency virus 1 (HIV-1), and herpes simplex virus type 1 (HSV-1), respectively, from the VirHostNet database~\citep{Navratil2009VirHostNet}. We used these sets of proteins for the analysis in \Cref{fig:prediction-results}(a).
}

\paragraph{Functional and protein interaction networks.} We used the human functional interaction network in the STRING database (version 11)~\citep{szklarczyk-mering-string-v10.5-nar-2017},
comprising of 18,886 nodes and 977,789 edges after applying a ``medium'' score cutoff of 400 and mapping to UniProt IDs. We used the interaction reliabilities provided by STRING as edge weights; we divided each value in STRING by $1,000$ to scale them between 0 and 1. An edge in this network may be derived from experimental data or computational analysis. Thus, an edge may represent either direct physical binding or indirect functional interaction. Of the 332 viral interactors, 328 were present in this network; REEP6 (Q96HR9), PPIL3 (Q9H2H8), RAB18 (Q9NP72), and FKBP7 (Q9Y680) were missing.

We also computed results for PPI networks from two other sources: the BioGRID database~\cite{biogrid-prot-sci-2021}, and the high-quality ``HI-union'' network published by Luck \emph{et al.}~\cite{luck-calderwood-ref-map-ppi-nature-2020}.
For BioGRID, we considered two versions: (i) all PPIs (including protein complex membership), and (ii) only direct PPIs from yeast two-hybrid (Y2H) screens.
For each of these networks, we did not use edge weights and restricted the nodes and edges to those in the largest connected component.
See \Cref{tab:net-stats} for statistics of the network size and density. 

\begin{table*}[hbt]\centering
    \centering
\begin{tabular}[c]{l|r|r|c|r|r|r}
Network & \# Nodes & \# Edges & Edge    & Density & \# \sarstwo & \# Nbrs.\\
        &          &          & Weights &         &  inter. (/ 333) & of sources\\
\hline
STRING (400) & 18,886 & 977,789 & Y & \(5.5\times{}10^{-3}\) & 329 & 12,480\\
BioGRID & 16,595 & 488,787 & N & \(3.6\times{}10^{-3}\) & 333 & 9,178\\
BioGRID-Y2H & 12,582 & 87,801 & N & \(1.1\times{}10^{-3}\) & 271 & 2,891\\
HI-union & 9,053 & 64,193 & N & \(1.6\times{}10^{-3}\) & 168 & 2,031\\
\end{tabular}
    \caption{Network statistics. 
    For STRING, the weight cutoff applied is in parentheses.
    The column titled ``\# \sarstwo inter. (/ 333)'' shows the number of sources that were in the network.
    The ``\# Nbrs. of sources'' column shows the number of neighbors of the human proteins that interact with \sarstwo proteins (i.e., sources) in the given network.}
    \label{tab:net-stats}
\end{table*}

\paragraph{Drug-protein interactions.} We downloaded interactions among drugs and proteins from the DrugBank database~(version 5.1.6)~\citep{2018-nar-wishart-wilson-drugbank-5}. This dataset contained 16,503 drug-protein target pairs among 5,665 drugs and 2,891 target proteins. Limiting the targets to those in the STRING network reduced the number of drugs and targets to 5,589 and 2,769, respectively.   

\paragraph{\sarstwo{}--human A549 AP-MS interactome.} 

We obtained 882 human proteins determined to interact with SARS-CoV-2 proteins by affinity purification followed by mass spectrometry
analysis (AP-MS) ~\citep{2020-biorxiv-stukalov-multi-level-proteomics}.  This dataset was generated in A549 lung carcinoma cells transduced with lentivirus vectors expressing HA-tagged SARS-CoV-2 proteins. The authors used affinity purification with anti-HA antibodies to isolate stable complexes of human proteins bound to SARS-CoV-2 proteins. Subsequently, they identified and quantified the purified proteins by mass spectrometry.  

\paragraph{\sarstwo{}--human HEK293 AP-MS interactome.} 

We obtained a set of 225 human proteins determined to interact with SARS-CoV-2 by AP-MS ~\citep{2020-biorxiv-li-guo-virus-host-interactome}.  This dataset was generated by analyzing HEK293 embryonic kidney cells transfected with plasmid vectors expressing FLAG-tagged SARS-CoV-2 proteins.  Affinity purification with anti-FLAG antibodies was used to isolate stable complexes of human proteins bound to SARS-CoV-2 proteins, and the purified proteins were identified and quantified by mass spectrometry.  

\paragraph{\sarstwo{}--human BioID interactome.} 
We obtained a set of 2,241 human proteins determined to interact transiently or weakly with SARS-CoV-2 proteins by using proximity-dependent biotinylation (BioID)~\citep{2020-biorxiv-samavarchi-tehrani-host-proximity-interactome}.  This dataset was generated by analyzing A549 lung carcinoma cells transduced with lentivirus vectors expressing SARS-CoV-2 proteins fused with a bacterial biotin ligase.  The addition of biotin resulted in the biotinylation of host proteins in the proximity of SARS-CoV-2 proteins.    Biotinylated proteins were purified and then identified and quantified by mass spectrometry.  Compared to interactomes identified by AP-MS, BioID is more capable of identifying weaker interactions in poorly soluble intracellular locations such as membranes and organelles.  

\paragraph{Differential protein abundance in \sarstwo{}--infected iAT2 cells.}
We obtained a set of 5,665 human proteins determined to have differential abundance in response to \sarstwo{} infection  ~\citep{2020-mol-cell-hekman-sars-cov-2-host-responses}.  This dataset was generated by infecting induced pluripotent stem cell-derived alveolar epithelial type 2 cells (iAT2) with \sarstwo{} and measuring protein abundance by quantitative mass spectrometry at 1, 3, 6, and 24 hours post-infection.  The authors compared protein abundance in infected iAT2 cells with that of the uninfected iAT2 controls to obtain differentially-expressed proteins. In our analysis, we used the set of proteins with differential expression (FDR \pval{} < 0.05) at any of the 1, 3, 6, and 24 hours post-infection.

\paragraph{Differential gene expression in upper airway samples in \sarstwo{}--infected patients.}
We obtained three sets of human proteins determined to have differential gene expression in cells infected with respiratory viruses~\citep{2020-nature-comm-mick-sars-cov-2}.  To generate this dataset, the authors used metagenomic RNA-seq to identify and quantify both human and viral RNA expression in upper airway samples collected from patients with acute respiratory illness.  They compared the gene expression values between samples that contained \sarstwo to uninfected samples in order to obtain differentially-expressed genes. They also identified additional viral infections including \sars, HRV, Influenza, HMPV, RSV, PIV in patient samples. Comparing \sarstwo infections with other viral infections and other viral infections with uninfected samples yielded two additional sets of differentially-expressed genes.  In our analysis, we used the genes with differential expression (FDR \pval{} < 0.05) in these three sets obtaining (i) 1,383 genes from \sarstwo-infected cells compared with uninfected samples, (ii) 7,338 genes from \sarstwo-infected cells compared with other viral infections, and (iii) 5,779 genes from other viral infections compared with uninfected samples.

From each of these interactome and differential expression datasets, we removed human proteins used as positive examples in our analysis and the proteins that were not present in the STRING network.  This step resulted in 2,080, 807, 212 proteins, respectively, from the interactome datasets and 5,447, 1,293, 6,940, and 5,472 proteins, respectively, from the differential expression datasets. We used Fisher's exact test to estimate the statistical significance of the overlap between the remaining proteins and our top-ranking proteins. 

\section{Analyses}
\label{sec:results}

Various network propagation methods have been successfully used in diverse applications in systems biology~\citep{cowen-sharan-network-propagation-amplifier-2017}. In particular, we model network propagation using the Regularized Laplacian (RL)~\citep{FOUSS201253}.  As we describe below (``Methods''), RL has the benefit of two mutually-reinforcing interpretations.  On one hand, it can be understood as an optimal labeling of network nodes, when some node labels are known \emph{a priori}. On the other hand, it can be seen as the result of diffusion, i.e., a continuous-time random walk, on the network. Under this second interpretation, we derived a novel mathematical formula for the expected length of the path traversed in the network by the random walker, which we then used to characterize our top-ranking proteins.

\subsection{Prioritization of Potential \sarstwo Interactors}
\label{sec:cross-validation}

Our underlying hypothesis was that network propagation via methods such as the RL yields a reasonable mechanism for predicting SARS-CoV-2 interactors. Therefore, we applied RL to the set of positive examples to rank the remaining proteins in the STRING network. We also ranked these proteins using multiple other network propagation methods and off-the-shelf classifiers~\citep{page-brin-pagerank-1999,mostafavi-morris-genemania-gb-2008,murali-katze-network-based-prediction-hiv-ploscb-2011,gligorijevic-bonneau-deepnf-bioinfo-2018}. We used a stratified sampling approach to estimate the statistical significance of the resulting node scores (see ``Statistical Significance of Node Scores'' in the supplementary methods). The sampling accounted for the possibility that if many sources have high degree, then scores may tend to be large overall in the network. Henceforth, for every method, we only considered proteins in the network that had a \pval less than 0.05.

To decide which methods to select for subsequent analyses, we compared them using 5-fold cross validation (``Comparison of Cross-Validation Results'' in the supplementary results and Figure S1). 
RL, random walk with restarts (RWR)\citep{page-brin-pagerank-1999}, and \deepnf~\citep{gligorijevic-bonneau-deepnf-bioinfo-2018} had the highest values of area under the precision-recall curve followed by SVM and logistic regression. RL achieved marginally worse values of area under the precision-recall curve than RWR and \deepnf. We selected one network propagation method (RL) and one supervised classifier (SVM) for the following reasons. We preferred RL over \deepnf because the provenance tracing method we developed for RL  enabled its results to be more easily interpreted than those for \deepnf. Since RL and RWR  produced highly similar predictions with a very high Spearman's correlation for the ranking of all proteins (``Overlap among algorithms'' in the supplementary results and Figure S2), we selected RL as representative of the two methods.
We chose SVM among the two off-the-shelf classifiers since it also had very good performance in cross-validation. 
We considered the top 332 predictions of RL and SVM that were statistically significant at $p < 0.05$~(``List of RL and SVM predictions, $p$-values, and top-two contributors''~\citep{law-murali-network-propagation-gigadb-2021}), which we refer to as ``top-ranking proteins'' below. 

Three recent publications or preprints have independently discovered physical interactions between \sarstwo and human proteins~\citep{2020-biorxiv-samavarchi-tehrani-host-proximity-interactome,2020-biorxiv-stukalov-multi-level-proteomics,2020-biorxiv-li-guo-virus-host-interactome}. These datasets differed in the type of host cell in which the viral proteins were expressed and the experimental methods used to determine if two proteins interacted. (``Datasets''). The top-ranking proteins for both RL and SVM had significant overlaps with each of the three new datasets, while the results for Local were not statistically significant (\pval $> 0.01$)~(\Cref{fig:prediction-results}(a)). We observed an especially striking overlap with the ``proximity interactome''~\citep{2020-biorxiv-samavarchi-tehrani-host-proximity-interactome}. Approximately one-third of the 332 and 1000 top-ranking proteins computed by RL were present in this dataset of 2,080 interactions (\pval $8.2 \times 10^{-24}$ and \pval $3.9 \times 10^{-74}$ respectively). 

The corresponding publication used BioID with the fast-acting miniTurbo enzyme~\citep{2020-biorxiv-samavarchi-tehrani-host-proximity-interactome}, a technique that is useful for discovering  viral-host protein interactions that take place at intracellular membranes and poorly soluble organelles, which are difficult to profile using classical biochemical purification approaches used in the other publications~\citep{2020-nature-gordon-krogan-sars-cov-2-human-ppis,2020-biorxiv-stukalov-multi-level-proteomics,2020-biorxiv-li-guo-virus-host-interactome}. Thus, our top-ranking proteins may be members of biological processes that occur in such locations in the cell. These three independent datasets provide strong support of our predictions. Our top-ranking proteins that do not overlap with these resources may interact with viral proteins indirectly and thus would not be captured by assays that test for direct protein-protein interactions.

We additionally tested the overlap between our top predictions and independent experimental datasets identifying differential expression of proteins in response to \sarstwo infection~\citep{2020-mol-cell-hekman-sars-cov-2-host-responses}. As in the previous analysis, we observed that the results for Local were not statistically significant (\pval $> 0.01$), while both RL and SVM had significant overlaps with differential protein abundance in \sarstwo infected cells compared with uninfected cells~\citep{2020-mol-cell-hekman-sars-cov-2-host-responses}(\Cref{fig:prediction-results}(a)).  Approximately half of the 332 and 1000 top-ranking proteins computed by RL were present in this dataset of 5,447 differentially expressed proteins (\pval $3.9 \times 10^{-14}$ and \pval $9 \times 10^{-62}$ respectively).   This high overlap may indicate that these proteins are involved with changes in host protein expression occurring in \sarstwo infected cells, via either direct or indirect virus-host protein interactions.

In contrast, when we analyzed gene expression measurements in response to \sarstwo{} infection~\citep{2020-nature-comm-mick-sars-cov-2}, we did not observe a significant overlap between our top-ranking proteins and differentially-expressed genes (Figure~S8). 
This result may be attributed to a difference in cell types used for measuring gene expression data, including cells not directly infected by the virus.  Moreover, the lack of edges connecting transcription factors to target genes in the PPI network we used may limit the size of the overlap between interactors predicted by RL and SVM with differentially-expressed genes.

We tested for enrichment of Gene Ontology (GO) biological processes (Benjamini-Hochberg corrected $p$-value $\leq 0.01$) among the top-ranking proteins from RL and from SVM, as well as in the interactors of SARS-CoV-2 (``Functional Enrichment'' in the supplementary methods).
Our top-ranking proteins were enriched in five broad categories of GO biological processes: organelle organization, transcription and translation, respiration, ER stress, and post-translational modifications (\Cref{fig:prediction-results}(b), Figure~S5
, and ``Enrichment results for RL, SVM and viral interactors''~\citep{law-murali-network-propagation-gigadb-2021}).
We examine the relevance of these processes to the viral life cycle in more detail in ``Discussion'' and in
``Enriched Biological Processes''
in the supplementary results.

\subsection{\new{Tracing the Provenance of Top-Ranking Proteins}} 


We can interpret the RL in terms of a continuous-time random walk over the network, which is governed by the internal parameter $\alpha$. We are interested in the node reached by the walker after a random time that depends on $\alpha$. The expected number of transitions made by the walker increases with the parameter $\alpha$ (``Analytical Perspective on the RL and Expected Path Length'' in the supplementary methods). Hence for larger values of $\alpha$, the ``influence'' of the sources is diffused more broadly across the network. To test how this spreading of ``influence'' affects our results, we varied $\alpha$ over four orders of magnitude from 0.01 to 100 and performed two analyses. First and most importantly, for each top-ranking protein computed by the RL, we developed a systematic procedure to determine the provenance of its score, i.e., which \sarstwo interactors made the greatest contributions to this score. For our second analysis, we developed a new methodology to select a value of $\alpha$. We were motivated to do so since we could not use the common practice of choosing the parameter's value based on maximization of cross-validation performance: the AUROC, AUPRC, and precision at 0.3 recall of the RL varied very little with $\alpha$ (Figure S3).

\begin{figure*}
    \centering
    \includegraphics[width=\textwidth]{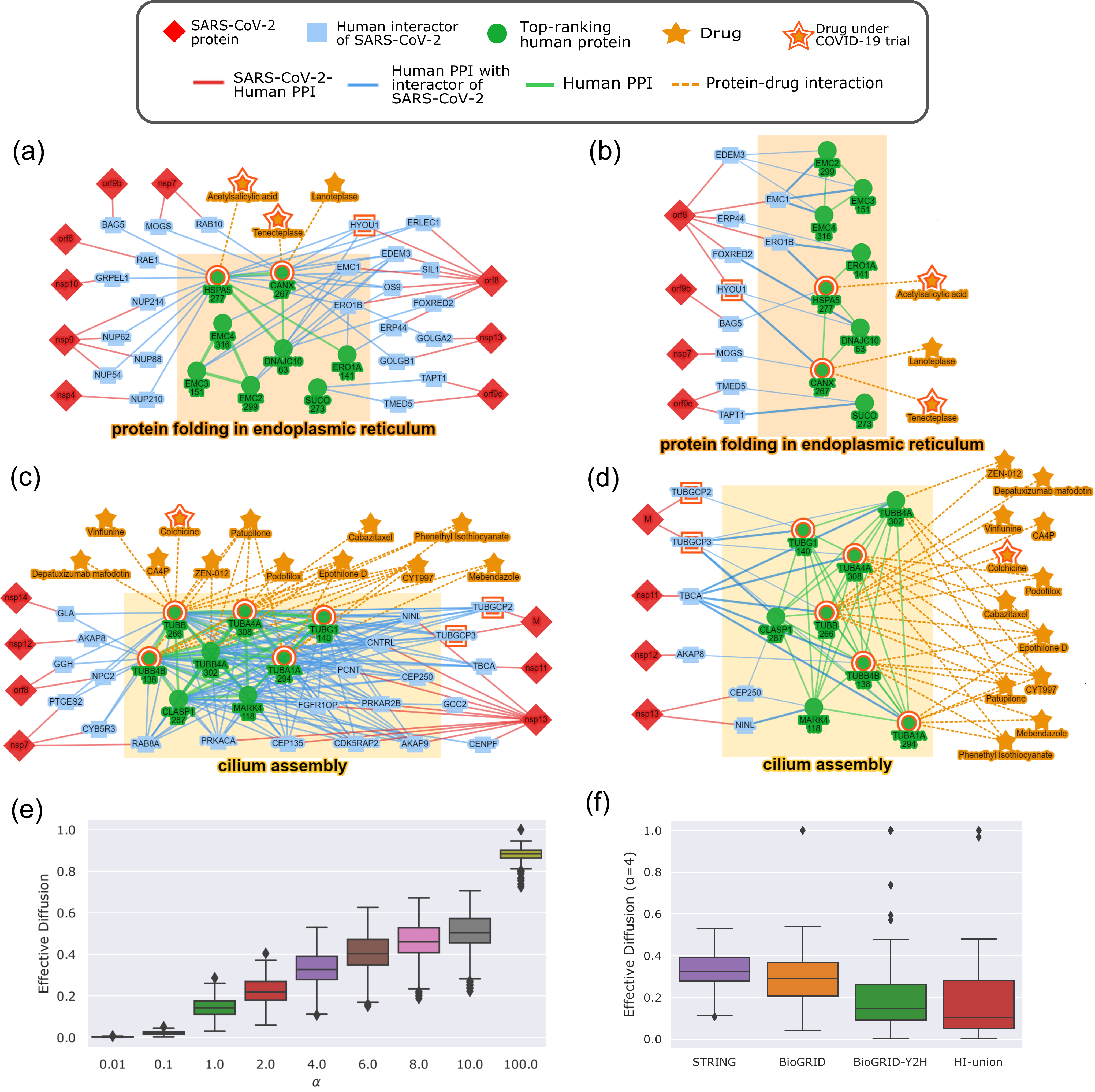}
    \caption{Provenance tracing results and illustrative examples of networks.
      (a) Network of the top 332 ranking proteins for RL (green nodes) that are annotated to the enriched term ``protein folding in ER''. For each top-ranking protein, we display its connections with all neighboring \sarstwo interactors.
      (b) The same network as in (a) except that we display only the top-two contributing \sarstwo interactors for each top-ranking protein.
      (c) Network of the top 332 ranking proteins for RL (green nodes) that are annotated to the enriched term ``cilium assembly.''
      (d) The same network as in (c) except that we display only the top-two contributing \sarstwo interactors for each top-ranking protein.     In all four network visualizations, the number below the name of a green protein is its rank as computed by the RL.
    Proteins discussed in the text are highlighted with a red border.
    In (a,c), we removed STRING edges with weight $< 700$ to simplify the visualization. We retained this restriction in (b,d) as well to maintain consistency between the visualized networks.
    In (c,d), we removed drugs that promote clotting. 
    (e) Distribution of effective diffusion for the top 332 ranking proteins for different values of $\alpha$.
    (f) The same distribution as (e) except comparing different networks with $\alpha = 4.0$.
    }
    
\label{fig:term-networks}
\end{figure*}

For provenance tracing, we took advantage of the fact that the score computed by the RL for each protein in the network is a linear combination of contributions from source proteins (``Methods''). Therefore, for each protein $u$ in the network, we sorted the source proteins by their relative contributions to the score of $u$ (``Provenance tracing matrix''~\citep{law-murali-network-propagation-gigadb-2021}). 
\Cref{fig:term-networks}(a)--(d) provide illustrative examples of the practical usefulness of provenance tracing. We used a value of $\alpha = 3.4$ to obtain these results. We present our method for selecting $\alpha$ at the end of this section.

In \Cref{fig:term-networks}(a), we display the top 332 ranking proteins computed by the RL that are annotated to the enriched GO term ``protein folding in endoplasmic reticulum''. For each such protein, we also show all the sources that interact with it as well as the viral proteins that in turn interact with the sources. This network is complex and difficult to understand. In contrast, in \Cref{fig:term-networks}(b), we connect each top-ranking protein only to the two source proteins that contribute the most to its score. This simplified network considerably facilitates the interpretation and rationalization of the RL's predictions. \Cref{fig:term-networks}(c,d) are similar in nature and correspond to the enriched term ``cilium assembly''. We return to the biological insights present in these networks in ``Discussion''.

Next, we considered the effect of $\alpha$ on the amount of diffusion in the network. 
When $\alpha$ was very small, e.g., $0.01$, we expected the highest contributing sources to be direct neighbors of top-ranking proteins. As $\alpha$ increased, and the random walker traversed longer paths in the network, we expected more of the highest contributors to not be directly connected by an edge to top-ranking proteins. Contrary to our expectations, we found that for every value of $\alpha$ and for every top-ranking protein $u$ (till a rank of 1,000), the source protein with the highest contribution to $u$'s score was always a neighbor of $u$. Even when we considered the second and third highest contributors, we found that they were more than one edge away for as few as 2\% of the top-ranking proteins for $\alpha=0.01$. This number increased only to 25\% for $\alpha=100$.

The STRING network includes both direct, physical and indirect, functional  PPIs. Therefore, we sought to see if this trend in the provenance analysis held for networks with only physical interactions corresponding to direct binding and indirect protein complex membership. We repeated the analyses up to this point on three other PPI networks: BioGRID, BioGRID-Y2H, and HI-union (``Methods''). 
For BioGRID, the results were comparable to those for STRING. The highest contributor was always a neighbor, except for $\alpha \geq 10$ where up to 3\% of nodes received most of their score from a source more than one edge away. The second and third highest contributor was more than one step away for as few as 8\% of top-ranking nodes for $\alpha=0.01$, and up to 41\% for $\alpha=100$.
For BioGRID-Y2H and HI-union, which are smaller, sparser networks with only direct PPIs, only 300--400 nodes had scores that were statistically significant at the 0.05 level. The highest contributing source was more than one step away for as many as 10--30\% of the top-ranking nodes, even for $\alpha=10$. For the second highest contributor, this percentage jumped to more than 50\% for $\alpha=0.01$ itself. 

To further characterize the contribution of non-neighboring sources, 
we defined the \emph{effective diffusion} to a protein $u$ as the fraction of its score $s(u)$ that arose from the non-direct neighbors of $u$ that were also \sarstwo interactors. As expected, the effective diffusion to the top-ranking proteins increased with $\alpha$ with values close to zero for $\alpha = 0.01$ and a median of 0.88 for $\alpha = 100$ (\Cref{fig:term-networks}(e)).
We concluded that the neighbours of the sources received non-trivial contributions to their RL scores from indirectly-connected sources only for values of $\alpha = 1$ and higher.

We repeated these experiments for BioGRID, BioGRID-Y2H, and HI-union (see \Cref{fig:term-networks}(f) and Figure~S9).
BioGRID maintained fairly similar results to STRING. On the other hand, for the other two networks, their effective diffusion values were quite a bit smaller (difference from STRING about 0.2 on average).
Taken together, these results suggest that in the sparser networks (BioGRID-Y2H and HI-union), a top-ranking protein has fewer sources as direct neighbours than in the denser networks (STRING and BioGRID) but a larger proportion of its score arises from these adjacent sources. 

These results motivated us to test a different method for selecting an appropriate value of $\alpha$ for downstream analysis. As mentioned earlier, we mathematically derived a new expression for the expected value of the path length of the random walker (``Analytical Perspective on the RL and Expected Path Length'' in the supplementary methods). To our knowledge, no such formula is known for the interpretation of the RL as a continuous-time Markov chain. This value depended on $\alpha$, the topology of the network, and which proteins interacted with \sarstwo. We computed the expected path length for different values of $\alpha$
(Table~S1).
Independently, we computed the distribution of path lengths in the network from \sarstwo{} interactors to every other protein (Figure~S10). 
The median number of edges in these paths was three. Therefore, we set the value of $\alpha = 3.4$ for which the expected path length of the random walker was 3.04
(Table~S1).
The median effective diffusion for this value of $\alpha$ was around 0.3. We used this value of $\alpha$ to generate the results presented in this work. 

\section{Discussion}
\label{sec:discussion}

The COVID-19 pandemic and its medical and economic impact have created an urgent challenge for biomedical researchers to understand infection mechanisms used by \sarstwo and to develop therapeutics against the disease~\citep{2020-science-guy-dutch-repurposing-covid-19}.  A manifestation of this community response
is the first protein-protein interactome associated with the \sarstwo-human interface~\citep{2020-nature-gordon-krogan-sars-cov-2-human-ppis}. This set of human proteins reported to interact with \sarstwo is likely to have both false positives and false negatives due to the properties of the proteomic screening pipeline used. 

In this work, we sought to further extend the results of this study to significantly expand the resources available to the \covid community by producing an extended set of putative \sarstwo interactors. Comparison of our results with independently-generated \sarstwo--human protein interaction networks~\citep{2020-biorxiv-samavarchi-tehrani-host-proximity-interactome,2020-biorxiv-li-guo-virus-host-interactome,2020-biorxiv-stukalov-multi-level-proteomics} provides substantial experimental support for our predictions. We note that complementary efforts are based on protein structures~\citep{
wu2020analysis}, observational studies of treatments being administered to patients~\citep{vaduganathan2020renin}, shortest paths in protein networks~\citep{zhou2020network}, propagation in protein networks with predicted \sarstwo interactors~\citep{2020-zhang-cai-sars-cov-2-rwr}, and exploratory analyses of virus-host-drug networks~\citep{2020-nat-comm-sadegh-baumbach-covex}. 

A notable new feature of our methodology is tracing the provenance of each of our predictions back to the most informative experimental sources~\citep{2020-plos-bio-kasif-roberts-reproducible-trace}. In principle, the RL computes scores by integrating over all paths in the network. We were surprised to see that the top-contributing sources were invariably direct neighbours of the top-ranking predictions in the STRING network. 
A partial explanation for this trend may be the fact that as many as 5,331 proteins in the STRING network were direct neighbors of at least one source protein, even when we considered only interactions with weight at least $0.9$ (the STRING database deems edges with such weights to be of ``very high quality''). Thus, the structure of the STRING network and central location of sources within it may cause the RL both to give high ranks only to direct neighbors of sources and to channel propagation primarily along these direct connections. We stress that using only the interactions between sources and their neighbors in the network does not result in high-quality predictions, as evidenced by the relatively poor cross-validation performance of the Local algorithm. Thus, the integration of multiple paths by the RL plays a key role in prioritizing which neighbors of the sources are more likely to be potential interactors of \sarstwo proteins than others.

COVID-19 research has focused disproportionately on a small set of human proteins~\citep{2020-elife-stoeger-amaral-covid-19-ignoring-host-genes}. Our research has the potential to expand the repertoire of host proteins that are studied in the context of COVID-19 and thereby open new directions of study of the disease. The cellular processes in which our top-ranking proteins participate suggest how the virus may infect human cells. We discuss two illustrative examples of the type of insights provided by our approach, highlighting several proteins targeted by drugs that are already in clinical trials for \covid. We remind the reader that we computed functions enriched in the top-ranking proteins, performed the provenance analysis independently, and then integrated the results in the protein networks we visualized.

\subsection{The Role of Endoplasmic Reticulum Stress, HSPA5, and Anti-Clotting Drugs}

Our analysis points to a connection among interactors of \sarstwo, proteins involved in endoplasmic reticulum (ER) stress, 
and anti-clotting drugs ((\Cref{fig:term-networks}(a,b))). 
%
The GO biological process ``protein folding in endoplasmic reticulum'' was enriched in the top-ranking proteins 
(\pval \pvalue{4.32}{-9} for RL 
and 0.28 for interactors of \sarstwo).
HSPA5, also referred to as glucose regulated protein (GRP78) or immunoglobulin binding protein (BiP) in the literature, is evolutionarily conserved from prokaryotes to humans~\citep{pmid24658275}. It has a repertoire of functions associated with ER stress response.
HSPA5 is usually localized in the ER. When the ER is stressed, HSPA5 can translocate to the cell surface, the nucleus and mitochondria~\citep{pmid20208072,pmid29654145}.
On the cell surface, HSPA5 plays a multi-functional role in cell proliferation, cell viability, apoptosis, and regulation of innate and adaptive immunity~\citep{pmid29654145,pmid21309747}.

HSPA5 has been proposed as a universal target for human diseases~\citep{pmid25546329}. It has increasingly well-documented essential interactions and activities during viral infections. In particular, 
the role of HSPA5 in viral entry and pathogenesis has been widely investigated. 
\sars{} infection has been shown to lead to ER stress and the up-regulation of HSPA5~\citep{pmid22028656,pmid16940539}. The S protein of \sars{} can induce transcriptional activation of HSPA5~\citep{pmid16940539}. This protein can serve as a point of attachment for both MERS-CoV and bat coronavirus (bCoV HKU9)~\citep{pmid29887526}. 
Both Zika virus and Japanese encephalitis virus use HSPA5 to prevent apoptosis and to help in viral replication~\citep{pmid25888736}. A recent molecular docking study has predicted HSPA5 as a potential receptor for the \sars S protein~\citep{pmid32169481}.
The observed expression \emph{in vitro}  of HSPA5 in airway epithelial cells suggests that it may serve as an additional receptor for \sarstwo{} in these cells~\citep{Aguiar2020.04.07.030742}. Based on our network-based analysis and support in the literature, we hypothesize that HSPA5 may serve as a co-receptor, a point of viral attachment, or aid in viral entry of \sarstwo.  

Blood hypercoagulability is reported to be common among COVID‐19 patients~\citep{pmid32282949}. 
Top-ranking proteins HSPA5 and CANX 
act as chaperones for pro-coagulant proteins such as Factor~V and Factor~VIII.  Once Factor~VIII is secreted, it binds to another pro-coagulant protein von Willebrand factor (vWF) to prevent degradation of clots~\citep{pmid9607108}. Although Factor~V, Factor~VIII, and vWF are not among the top-ranking proteins and thus do not appear in \Cref{fig:term-networks}(a,b), this network is suggestive of mechanisms that \sarstwo may use to cause abnormal blood coagulation. 

Anti-coagulant drugs that interact with  HSPA5 or CANX 
include Tenecteplase, a third generation plasminogen activating enzyme and the investigational drug  Lanoteplase, which is a serine protease that binds to fibrin leading to the formation of plasmin~\citep{pmid22775207}, an enzyme that breaks clots. Lanoteplase is a second-generation derivative of Alteplase, 
and a third generation derivative of recombinant plasminogen. It is notable that there are clinical trials for Tenecteplase (ClinicalTrials.gov, NCT04558125, NCT04505592) and Alteplase (ClinicalTrials.gov, NCT04357730, NCT04640194) to test their effectiveness in treating COVID-19.  
Aspirin, also present in (\Cref{fig:term-networks}(a,b)), binds to and inhibits the ATPase activity of HSPA5~\citep{pmid11689471}. Aspirin is currently involved in 16 clinical trials (ClinicalTrials.gov), with one testing the effects of aspirin at various levels of COVID-19 severity (NCT04365309), and another testing whether early treatment of COVID-19 patients with aspirin and vitamin D can inhibit the production of blood clots and decrease rates of hospitalization (NCT04363840).

\subsection{Cilium Assembly and Tubulin-Modulating Drugs}  

GO biological processes related to cilia were significantly enriched in the top-ranking RL and SVM predictions.
An example is ``cilium assembly'' ($p$-value \pvalue{6.84}{-26} for RL vs. 
$0.31$ in the human interactors of SARS-CoV-2.
Many proteins annotated to this term belong to the tubulin family, which are components of microtubules. The SARS-CoV-2 M protein binds to two $\gamma$-tubulins (TUBGCP2 and TUBGCP3), which interact with several $\alpha$- and $\beta$-tubulins among the top 332 predictions 
(\Cref{fig:term-networks}(c,d)).
Microtubules are polymers that provide shape and structure to eukaryotic cells and are necessary in cell transport and cell division, among other functions~\citep{2000-nogales-annu-rev-biochem-microtubule}. $\alpha$- and $\beta$-tubulins compose microtubule filaments, while $\gamma$-tubulins connect them to the microtubule organizing center.  

Viruses commonly utilize microtubules for cellular entry, intra-cellular trafficking, and exit from cells~\citep{pmid16497585}. For instance, the S protein of human $\alpha$-coronavirus interacts with tubulin $\alpha$ and $\beta$ chains~\citep{2016-virology-rudiger-wessels-tubulins-s-proteins}, suggesting that tubulin may be involved in the transport and localization of the S protein and its assembly into virions~\citep{2016-virology-rudiger-wessels-tubulins-s-proteins}. Relevant to SARS-CoV-2, microtubules are the primary structural component of cilia, which line epithelial cells in the respiratory tract and are responsible for the transport of mucus out of cells~\citep{2007-annu-rev-physiol-satir-christensen-mammalian-cilia}. The ACE2 receptor that SARS-CoV-2 uses to enter cells appears to be expressed primarily on the cilia of respiratory tract epithelial cells~\citep{2020-nature-comm-lee-jackson-robust-ace2-expression-cilia,2020-nat-med-sungnak-figueiredo-sars-cov-2-entry-factors}, further implicating microtubules in viral infection. The combination of high expression levels of ACE2 and the presence of cilia may also explain the detection of the virus in multiple organs~\citep{2020-puelles-huber-nejm-multiorgan-renal-sars-cov-2} and the deleterious effect of COVID-19 on the renal, gastroinstestinal, and olfactory systems~\citep{pmid31986264}. The drugs that target Tubulin proteins (\Cref{fig:term-networks}(c,d)) are mostly anti-mitotic agents, which are being investigated as anti-cancer therapeutics. 
It is notable that 26 ongoing clinical trials (ClinicalTrials.gov) are testing the effectiveness of Colchicine against \covid. 

Our work also sets the stage for follow-up analyses on \sarstwo. Integrating new datasets of \sarstwo-human protein interactions~\citep{2020-biorxiv-samavarchi-tehrani-host-proximity-interactome,2020-biorxiv-li-guo-virus-host-interactome,2020-biorxiv-stukalov-multi-level-proteomics} and human proteins whose deletion inhibits viral replication \citep{2020-biorxiv-wei-alfajaro-genome-wide-crispr-screen,2020-cell-daniloski-host-factors-for-sars-cov-2}
with other omics data using our methods and with orthogonal analysis techniques promises to predict more biologically meaningful networks and processes impacted by the virus. In particular, single-cell RNA-seq data offer many opportunities to examine cellular heterogeneity and context-specific interactions.

\section{Potential Implications}
\label{sec:potential-implications}

The approach we advocate here is inspired by the general framework of producing explanations for machine learning methods~\cite{2016-kdd-ribeiro-guestrin-explaining-classifier}. This area of ``explanations'' of predictions is receiving strong interest because of deep learning. While the idea has previously been studied in graphical models~\cite{2009-pearl-causality},  most machine learning methods are not fully interpretable by the fairly strict definition of Kasif and Roberts~\cite{2020-plos-bio-kasif-roberts-reproducible-trace}: tracing each prediction to the experimental evidence that supports it. This notion of explanation is a special but particularly important case for computational genomics and systems biology.

Causal perturbations~~\cite{2009-pearl-causality} provide a general approach for producing explanations of this type for virtually any predictive model. Consider a model with experimental evidence that a gene $g$ performs a function $f$. We perturb the variable associated with the gene, e.g. we change the probability $\Pr(\text{$g$ performs $f$}) = 1$ to $\Pr(\text{$g$ performs $f$}) = 0$.
We then compute the change in probability of every other variable in the model due to this perturbation in order to assess the importance of this particular gene-function pair.

For network propagation, this idea yields the special case discussed in this work that is amenable to very efficient computation. 
Our strategy for tracing provenance extends to any algorithm that makes predictions using a linear combination of evidence such as logistic regression and GeneMania~\citep{MRW+08}. In particular, it is applicable to the large number of random-walk-based methods that have been developed for predicting disease genes or annotations to GO terms~\cite{vanunu-sharan-associating-genes-and-complexes-with-disease-ploscb-2010,KWR10,Jiang-Gribskov-AptRank-protein-function-prediction-bioinfo-2017,2020-cell-syst-hristov-singh-ukin}.

An important future line of research will be to develop provenance tracing techniques for other classes of network-based methods such as Markov random fields (MRFs)~\cite{Letovsky2003Predicting,DTS+04} and min-cut based methods~\cite{mwk-agfp-2006,nabieva-singh-function-prediction-via-interaction-maps-2005}. For MRFs, we can apply the general perturbation-based method described above. For mincut-based methods, it is possible to recalculate the cut for any single change in experimental data using dynamic data structures~\cite{2018-talg-goranci-thorup-incremental-min-cut}. Thus, the provenance tracing approach that we advocate here has many natural follow-ups that we expect to be studied by the community in the future.

It remains to be seen whether the trends we observed on the contributions from direct neighbors generalize to these methods and to annotations of terms in the Gene Ontology or the Human Phenotype Ontology terms. In general, it is quite likely that sources that are not direct neighbors may make substantial contributions to scores. In these cases, new algorithmic developments may be required to trace the paths by which the sources spread their influence to a given node.

Our work provides significant new data and software resources to the \covid community. Three properties of our results facilitate their use by experimentalists who are seeking to obtain new insights into the pathogenesis of this disease. First, the prioritized list of predicted interactors of \sarstwo (``List of RL and SVM predictions, $p$-values, and top-two contributors''~\citep{law-murali-network-propagation-gigadb-2021}) contains druggable targets that may be promising to study further. Second, our provenance analysis provides the rationale underlying each prediction by directly linking to the relevant experimental input. Third, the viral-human protein interaction networks corresponding to enriched GO terms (\Cref{fig:term-networks} and Figure~S6) 
are available for visualization and download on GraphSpace (\url{http://graphspace.org/graphs/?query=tags:2021-sarscov2-network-analysis}). Examination of these networks provides further context for the predictions.

We conclude by noting that our methodology is general purpose and easy to generalise to a new virus. The software requires a dataset of host proteins that interact with the virus and an interaction network among the host proteins themselves. The virus-host network may be determined experimentally~\citep{2020-nature-gordon-krogan-sars-cov-2-human-ppis}. If such a dataset is not available, a user can predict the network computationally from the sequence of the viral genes and interaction networks for phylogenetically similar viruses~\citep{2021-psb-kshirsagar-klein-seetharaman-protein-sequence-models}. Subsequently, a user can apply network propagation to predict additional human proteins and biological processes that may be targeted by the virus. 

\section{Methods}
\label{sec:methods}

\subsection{Algorithms}
\label{sec:algorithms}

To facilitate the complete reproducibility of our results, we now describe the RL algorithm that we use for label propagation and prediction. We present the other methods that we use (\genemania, \sinksource, \rwr, \local, \deepnf, the Support Vector Machine, and Logistic Regression) and implementation details in ``Other Algorithms'' in the supplementary methods. We are given a weighted, undirected network $G = (V,E,w)$, where each node in $V$ is a human protein, each edge $(u, v)$ represents an interaction between proteins $u$ and $v$, and $w: E \rightarrow (0, 1]$ is a function specifying the weight of each edge in $E$. Informally, the weight of an edge indicates our confidence in the experimental data supporting the corresponding protein-protein interaction. We are also given a set $P \in V$ of positive examples consisting of the human proteins that interact with \sarstwo  proteins~\citep{2020-nature-gordon-krogan-sars-cov-2-human-ppis}. Each node in $G$ is a human protein and each edge represents a physical or functional interaction between two proteins. 
We seek to compute a score vector $\vec{s} \in \reals^n$, where $n$ is the number of nodes in $G$. For every node $v$, the score $s(v)$ in this vector  indicates our confidence that node $v$ either physically interacts with or is functionally linked to a \sarstwo protein.

\paragraph{Regularized Laplacian~\citep{FOUSS201253}.}
Given a parameter $\alpha > 0$, we compute $\vec{s}$ using the following steps:
    \begin{enumerate}
        \item Define a label vector $\vec{y}$ over the nodes in $G$ where $y(u) = 1$ if node $u$ is in $P$ and $y(u) = 0$, otherwise.
        \item Define $W \in \reals^{n\times{}n}$ as the adjacency matrix of $G$ with edge weights, i.e., the entry in row $u$ and column $v$ of $W$ equals $w_{uv}$ if $(u, v)$ is an edge in $G$ and $0$, otherwise. 
        \item Define $D$ as a diagonal matrix with $D_{uu} = \sum_v w_{uv}$, for every node $u$ in $G$.
        \item Compute the $\reals^{n\times{}n}$ matrix $\tilde{W} = D^{-1/2} W D^{-1/2}$, which denotes the normalized network.
        \item Compute the Laplacian of $G$ as $\tilde{L} = \tilde{D} - \tilde{W}$, where
        we define $\tilde{D}$ to be a diagonal matrix with $\tilde{D}_{uu} = \sum_v \tilde{w}_{uv}$.
        \item Compute the vector $\vec{s} = (I + \alpha\tilde{L})^{-1} \vec{y}$.
    \end{enumerate}

The RL was introduced by Zhou and Sch\"olkopf. Since then, several variations of this method have been published. The version we use is identical to the strategy used by Fouss \emph{et al.}~\citep{FOUSS201253}. We provide the intuition behind the resulting RL matrix (i.e., $(I + \alpha \tilde{L})^{-1}$) and discuss its properties in
``Analytical Perspective on the RL and Expected Path Length'' in the supplementary methods. In particular, we derive an expression for the expected path length of the continuous-time Markov chain corresponding to the RL. As far as we know, this mathematical analysis has not previously been published.

\subsection{Tracing the Provenance of Prediction Scores}


Let $K$ denote the RL matrix $(I + \alpha\tilde{L})^{-1}$.
We remind the reader that the RL algorithm ranks proteins based on diffusion scores that associate a node $u$ in the network with a diffusion score $s(u)$, where  $s(u) = \sum_{v\in P} K_{uv}$, where $v$ ranges over the set $P$ of all SARS-CoV-2 interactors.  For every protein $u$, we sorted the proteins in $P$ in decreasing order of the values of $K_{uv}$, where $v$ ranged over $P$. In the manner, we ranked the experimentally determined interactors that in decreasing order of their contributions to each node's diffusion score. This analysis is important for tracing the provenance of computational predictions to their experimental sources~\citep{2020-plos-bio-kasif-roberts-reproducible-trace}. 

\section{Availability of Source Code and Requirements}

\begin{itemize}
\item Project name: SARS-CoV-2-network-analysis
\item Project home page: \url{https://github.com/Murali-group/SARS-CoV-2-network-analysis}
\item Operating system(s): Platform independent  (tested and applied on Linux and Mac OS)
\item Programming language: Python
\item Other requirements: Please see \url{https://github.com/Murali-group/SARS-CoV-2-network-analysis/blob/master/requirements.txt}
\item License:~GNU General Public License (GPL) v3
\item biotools id: biotools:sars-cov-2-network-analysis
\item SciCrunch Resource id: SCR\_021811 
\end{itemize}

\section{Availability of supporting data and materials}

We used publicly available datasets for our analysis. We downloaded these data from the respective publications or websites. A snapshot of the software used for this analysis and the following supplementary files are available at the GigaScience GigaDB database~\citep{law-murali-network-propagation-gigadb-2021}.


\begin{description}
    \item[List of RL and SVM predictions, $p$-values, and top-two contributors:] The prediction rank and $p$-value computed by RL and SVM for each human protein on the STRING network, the list of drugs that target the protein (when this information is available in DrugBank), and the top-two contributing SARS-CoV-2 interactors and corresponding SARS-CoV-2 protein. For the last piece of information, we also included the fraction of score contributed by each of the top-two SARS-CoV-2 interactors.
    \item[Enrichment results for RL, SVM and viral interactors:] Enrichment results for RL, SVM and the viral interactors on GO biological processes. 
    \item[Provenance tracing matrix:] Provenance tracing matrix of contributions to the network propagation score from each \sarstwo interactor to every top-ranking protein. 
\end{description}

\section{Declarations}

\subsection{Abbreviations}

AP-MS: affinity purification followed by mass spectrometry analysis; AUPRC: area under the precision-recall curve; AUROC: area under the receiver-operator characteristic curve; bCoV:  bat coronavirus; COVID-19: novel coronavirus disease 2019; BioID: proximity-dependent biotinylation; BiP: immunoglobulin binding protein; ER: endoplasmic reticulum; GM: GeneMania; GPL: General Public License; GO: Gene Ontology; GRP: glucose regulated protein; HIV-1: human immunodeficiency virus 1; HSV-1: herpes simplex virus type 1; KEGG: Kyoto Encyclopedia of Genes and Genomes; LogReg: Logistic Regression; MERS: Middle East respiratory syndrome; NSF: National Science Foundation; PPMI: Positive Pointwise Mutual Information; RWR: random walk with restarts; RL: Regularized Laplacian; SARS: severe acute respiratory syndrome; SARS-CoV-2: severe acute respiratory syndrome coronavirus 2; SS: SinkSource; SVM: Support Vector Machine; vWF: von Willebrand factor; USDA-NIFA: United States Department of Agriculture National Institute of Food and Agriculture

\subsection{Consent for Publication}
Not applicable.

\subsection{Competing Interests}

The authors declare that they have no competing interests.

\subsection{Funding}

TMM acknowledges support from National Science Foundation (NSF) grants DBI-1759858 and MCB-1817736. KA acknowledges support from the Genetics, Bioinformatics, and Computational Biology program at Virginia Tech. JK acknowledges support from NSF grant CCF-2029543. MC acknowledges support from NSF grant CNS-1618207.  CMDS acknowledges support from the Hariri Institute and the Department of Biomedical Engineering at Boston University. PR acknowledges support from NSF grant CBET-1510920 and USDA-NIFA grant 2018-07578. PR and TMM  acknowledge support from the Computational Tissue Engineering Graduate Education Program at Virginia Tech.

\subsection{Author Contributions}

TMM and SK proposed the study. TMM, SK, MC, JL, SD, MK, and JK contributed computational ideas. JL was the primary author of the software and led the computational analysis, with significant inputs from KA, NT, and CMDS. All authors analyzed the results. TMM, MC, PR, and SK wrote the paper with contributions and revisions from all authors. All the authors read and approved the final manuscript.

\subsection{Acknowledgments}

The authors wish to thank S. Alabdullatif, S. Alshuaib, M. Iennaco, M. Kouzminov, S. Murthy, S. Makwana, N. Naguib, C. Tagliettii, and M. Zanna for exploratory research on this data and insightful and thought-provoking analysis.  We also thank Roded Sharan, Noga Alon, Dan Lancour and Rich Roberts for discussions that helped formulate the techniques and ideas we used in this paper. 

\bibliography{csb,Mendeley-Murali-group-references,meghana-refs}

\end{document}